\documentclass{PoS}

\title{Double Higgs production measurements from $HH\to(b\bar{b})(b\bar{b})$ at a 100 TeV hadron collider}

\ShortTitle{Double Higgs production from $HH\to(b\bar{b})(b\bar{b})$ at a 100 TeV hadron collider}

\author{\speaker{Nathan P. Hartland}\\
Rudolf Peierls Centre for Theoretical Physics,\\
1 Keble Road, University of Oxford, OX1 3NP, Oxford, UK.\\
\email{nathan.hartland@physics.ox.ac.uk}}

\abstract{In this contribution we study the prospects of measuring double Higgs production at a potential 100 TeV future circular collider. We apply an analysis procedure that utilises reconstructed Higgs pairs
from multiple final state event topologies in order to maintain high selection efficiencies. Signal purity is then further improved by means of a artificial neural network classifier. The results of this analysis for the high luminosity LHC show significant potential, however when applied to a 100 TeV hadron collider we find that such a measurement is likely to suffer from a very poor signal to background ratio. Such a measurement at the FCC is therefore likely to be significantly more challenging than at the high luminosity phase of the LHC.}

\FullConference{XXIV International Workshop on Deep-Inelastic Scattering and Related Subjects\\
		11-15 April, 2016\\
		DESY Hamburg, Germany}

\begin{document}

\section{Introduction}

\begin{figure}[h]
\begin{center}
\includegraphics[width=0.9\textwidth]{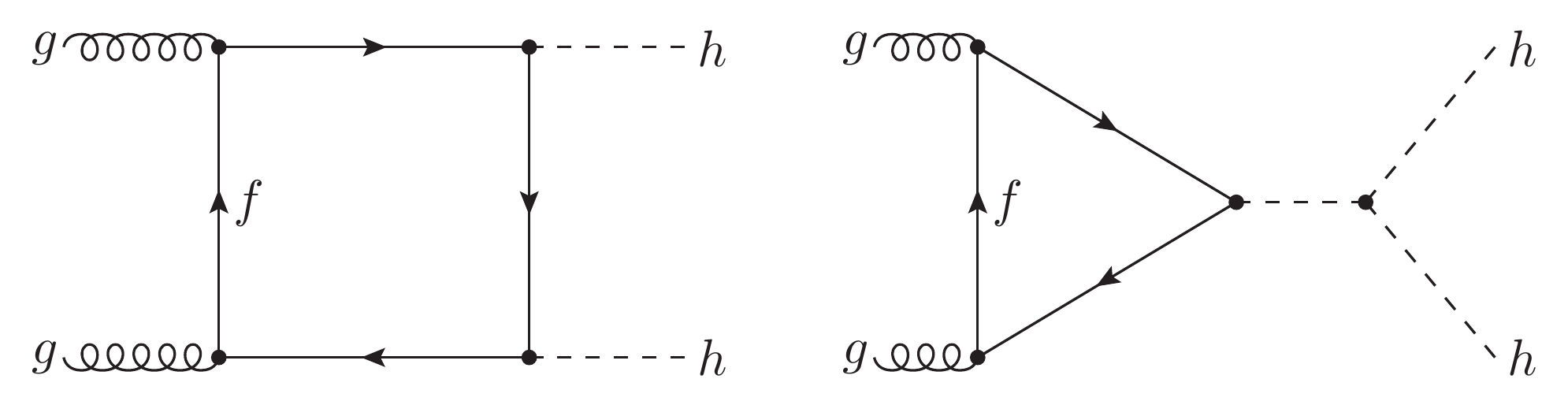}
\caption{\small Leading order diagrams for double Higgs production in gluon fusion.}
\label{fig:feyn}
\end{center}
\end{figure}

Measurements of Higgs boson properties at the Large Hadron Collider (LHC) covering a wide range of production channels and final states are incrementally building a picture consistent with our expectations for a Standard Model (SM) Higgs boson. The success of the LHC therefore revolutionising our understanding of the electroweak symmetry breaking mechanism. Nonetheless, several parameters of the Higgs mechanism remain unprobed.
Experimental results to date have shed light upon single Higgs production, and so have explored the minimum of the electroweak symmetry breaking potential. To understand the full potential, and therefore to probe several possibilities for potential Beyond the SM (BSM) physics, an observation of double Higgs production is a necessity. However the measurement of such a process at the LHC is a singularly challenging task, principally due to very low production rates. Di-Higgs systems at the LHC arise predominantly through the gluon-gluon fusion (ggF) subprocesses, analogously to the case of a single Higgs. The diagrams contributing to this subprocess are illustrated to leading order in Figure~\ref{fig:feyn}. The expected SM production rates for gluon fusion at 14 TeV are around a thousand times lower than those of single ggF Higgs production, with a NNLO+NNLL SM cross-section of around 40fb~\cite{deFlorian:2015moa}. Other HH production channels such as vector boson fusion or associated production with top pairs or electroweak gauge bosons have SM cross-sections of order $1$ fb. Compounding the already small production cross-sections, the bulk of the cross-section decays to the experimentally challenging fully hadronic final state. The combination of dominant production channel (ggF) and dominant decay mode ($b\bar{b}b\bar{b}$) therefore resulting in one of the more experimentally challenging systems to reconstruct.

Several groups have investigated the feasibility of such a measurement at the LHC and at its proposed high luminosity phase (HL-LHC) (See \cite{Behr:2015oqq} and references therein). Despite considerable differences in analysis procedures, these studies report that statistical signal significances of between one and three standard deviations may be possible at the HL-LHC. With recent discussions on the experimental potential of a possible future 100 TeV circular hadron collider (FCC100) the question of how such analyses would fare in the FCC environment merits investigation.

In this contribution we shall apply an analysis strategy recently developed for the detection of double Higgs production at the LHC~\cite{Behr:2015oqq} to event samples at the proposed FCC centre of mass energy of 100 TeV in order to assess the feasibility of such a measurement at the proposed machine.

\section{Measuring $HH\to(b\bar{b})(b\bar{b})$ at hadron colliders}

In \cite{Behr:2015oqq} we have proposed a procedure for the reconstruction of Higgs candidates across a wide range of Higgs transverse momenta. Typical measurement strategies
attempt to reconstruct the Higgs decay products either as a resolved pair of $b$-jets with small $R$ parameter, or by merging the decay products into a single large-$R$ jet. Naturally the efficiency of these reconstruction procedures varies with the collimation of the decay products and therefore with the boost of the parent Higgs. In order to retain as much as possible of the already small production cross-section, we propose a combination of three reconstruction topologies: \emph{fully resolved} constructed by forming di-jet pairs from a system of four small-$R$ b-jets, \emph{intermediate} where the decay products of one Higgs are reconstructed in a large-$R$ jet, and a pair of small-$R$ jets resolve the second Higgs decay products. Finally the \emph{fully boosted} channel reconstructs both Higgs bosons in merged, large-$R$ jets. Such a strategy has also been proposed as a tagging method that can likewise reconstruct heavy resonances across a wide range of scales~\cite{Gouzevitch:2013qca}.

In our analysis strategy, resolved $b$-jets were reconstructed as anti-$k_T$~\cite{Cacciari:2008gp} $R=0.4$ jets with $p_T\ge 40$ GeV and $|\eta| < 2.5 $. Merged Higgs decay products were reconstructed with Mass-Drop Tagged~\cite{Butterworth:2008iy} anti-$k_T$, $R=1.0$  jets with $p_T > 200 $ GeV and $|\eta| < 2.0$. The tagging of $b$-jets was simulated by requiring a $b$ quark with $p_T \ge 15$ GeV as a jet constituent. Tagging has a simulated efficiency of 80\%, a light-jet mistag rate of 1\% and a charm mistag rate of 10\%. 

Higgs candidates are therefore reconstructed as either a boosted Higgs consisting of a double $b$-tagged large-$R$ jet or as a pair of small-$R$ $b$-jets. In the fully resolved topology, dijet pairs are chosen by selecting the four hardest small-R jets in the event, and choosing the pairings that minimise the Higgs candidate mass difference $|m_{H1} - m_{H2}|$. All Higgs candidates must furthermore pass a mass window cut of $40$ GeV around $m_H=125$ GeV. Naturally in this configuration there is some overlap between the event topologies, this is broken by applying a strict ordering upon the categories. Firstly our analysis attempts to categorise each event into the boosted topology. If the event fails the boosted kinematic cuts, the analysis attempts to reconstruct the event within the intermediate category. If this again fails the analysis finally attempts to reconstruct the event with the resolved topology.

Following this cut-based analysis, surviving events are categorised as either signal or background by a trained artificial neural network performing a multivariate analysis (MVA). Each analysis topology has an independent neural network trained to provide a useful discriminant, given an input set of kinematic variables characterising each event. In the case of the intermediate and boosted topologies we make use of the discrimination power of jet substructure techniques in the MVA. In addition to standard kinematics, for each large-R jet the neural networks receive as input the 2-to-1 subjettiness ratio~\cite{Thaler:2010tr}, $k_T$ splitting scales~\cite{Butterworth:2002tt}, and the energy correlation function double ratios $C_2$~\cite{Larkoski:2013eya} and $D_2$~\cite{Larkoski:2014gra}.
The initial cut-based analysis strategy has been kept deliberately loose in order to maximise the amount of information available to the MVA, the output of which is a further real-valued discriminant which may be cut upon to improve signal significance and purity.

\subsection{Simulation}
Signal samples were generated at leading order (LO) with {\tt Madgraph5\_aMC@NLO}~\cite{Hirschi:2015iia} in the four-flavour scheme, with $\alpha_S=0.118$ and a Higgs mass of $m_H=125$ GeV.  NNPDF3.0 LO PDFs~\cite{Ball:2014uwa} with the corresponding settings were used. Factorisation and renormalisation scales were set at $\mu_F=\mu_R=H_T/2$. The signal sample is showered with {\tt Pythia} 8.201~\cite{Sjostrand:2014zea}. For the background, while previously we included all relevant processes in the analysis, here we shall consider only the irreducible QCD $4b$ background for comparison between the HL-LHC and FCC. The QCD 4b backgrounds have been generated with {\tt Sherpa} 2.1.1~\cite{Gleisberg:2008ta} with the same parameters as for the signal generation. In these comparisons no pileup effects are included.
\section{Results}

The results of running signal and QCD $4b$ samples at 14 and 100 TeV through the cut-based analysis chain are shown in Table~\ref{tab:postanalysis}. Here we have assumed an integrated luminosity of 3 ab$^{-1}$ for the HL-LHC and 10 ab$^{-1}$ for the FCC. While all cross-sections naturally grow between the HL-LHC and FCC configurations due to the increased luminosity, the intermediate and boosted channels demonstrate considerably larger growth due to the enlarged $p_T$ reach available with the higher centre of mass energy. Thanks to the higher collider energy and target luminosity, the FCC demonstrates considerably greater statistical significances than those of the HL-LHC, reaching a combined statistical significance of around 7.5 standard deviations after only the cut-based analysis. However the signal over background ratio notably deteriorates in the FCC configuration due to the greater increase in the background when moving between the two collider environments. While at the HL-LHC, $S/B$ ratios at the permille level were achievable at the level of the cut-based analysis, at the FCC this does not appear to be possible even when only considering the QCD 4b background.

\begin{table}[htp]
\begin{center}
\begin{tabular}{|l|c|c|c|c|}
\hline
\multicolumn{5}{|c|}{14 TeV HL-LHC (3 ab$^{-1}$)}\\
\hline
\hline
Topology & $HH$ (fb) & $4b$ (fb) & $S/\sqrt{B}$ & $S/B$ \\
\hline
Resolved      &0.5	& $1.7\times10^3$ & 0.6 & $2.9\times10^{-4}$ \\
Intermediate &0.09	& $5.6\times10^1$ & 0.6 & $1.6\times10^{-3}$ \\
Boosted        &0.16	& $5.3\times10^1$ & 1.1 & $2.7\times10^{-3}$\\
\hline
\end{tabular}
\vskip10pt
\begin{tabular}{|l|c|c|c|c|}
\hline
\multicolumn{5}{|c|}{100 TeV FCC (10 ab$^{-1}$)}\\
\hline
\hline
Topology & $HH$ (fb) & $4b$ (fb) & $S/\sqrt{B}$ & $S/B$ \\
\hline
Resolved      &11.4	& $7.8\times10^4$ & 4.1 & $1.5\times10^{-4}$ \\
Intermediate &3.1	& $9.9\times10^3$ & 3.1 & $3.1\times10^{-4}$ \\
Boosted        &4.9	& $7.5\times10^3$ & 5.7 & $6.5\times10^{-4}$\\
\hline
\end{tabular}
\label{tab:postanalysis}

\caption{Results of the cut-based analysis for 14 TeV HL-LHC (upper table) and 100 TeV FCC (lower table). For each collider setup and analysis topology, cross-section yields for the signal
and background process are provided along with the corresponding signal significance and purity.}
\end{center}
\end{table}%

\begin{figure}[t]
\begin{center}
\includegraphics[width=0.45\textwidth]{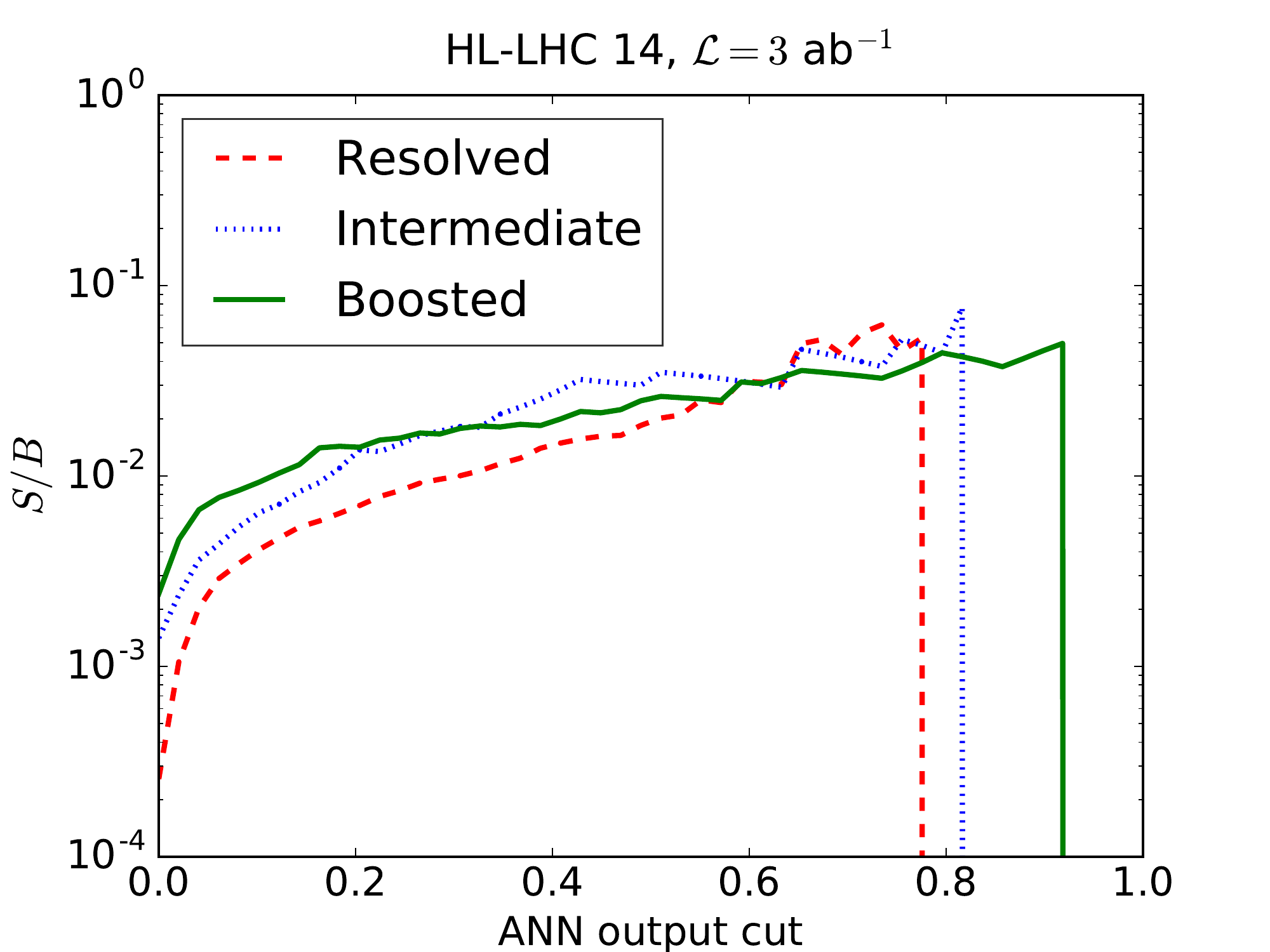}
\includegraphics[width=0.45\textwidth]{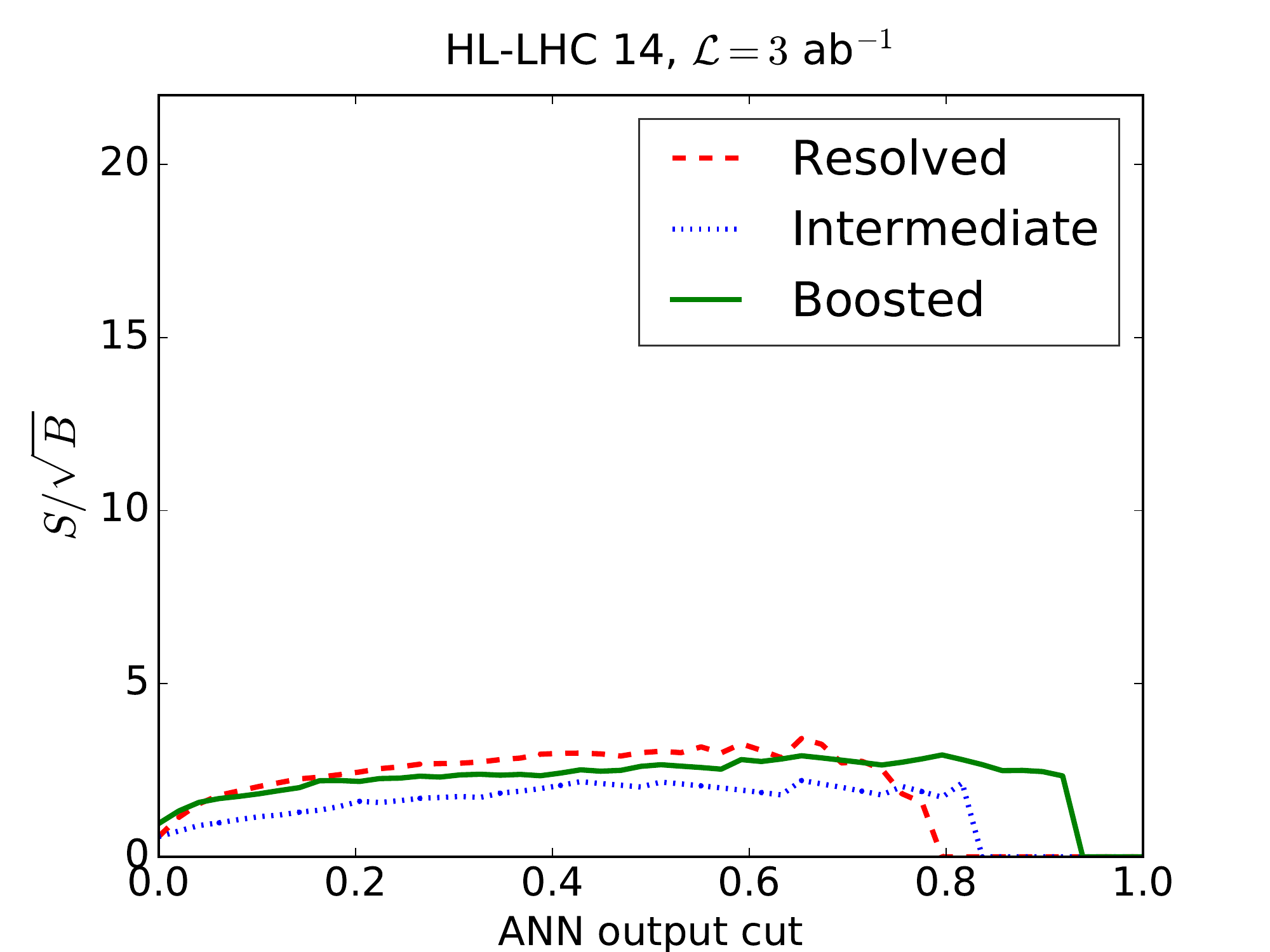}\\
\includegraphics[width=0.45\textwidth]{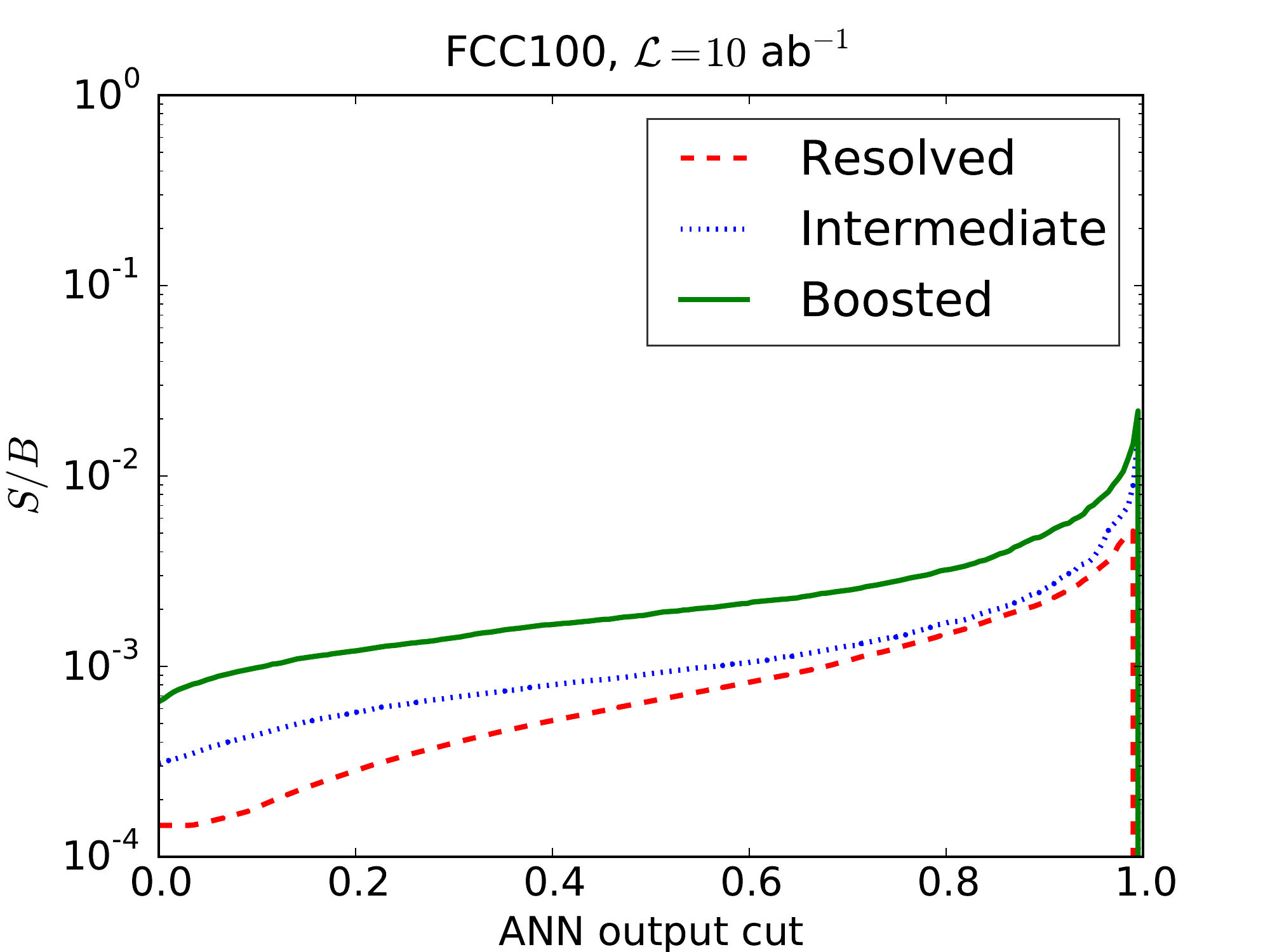}
\includegraphics[width=0.45\textwidth]{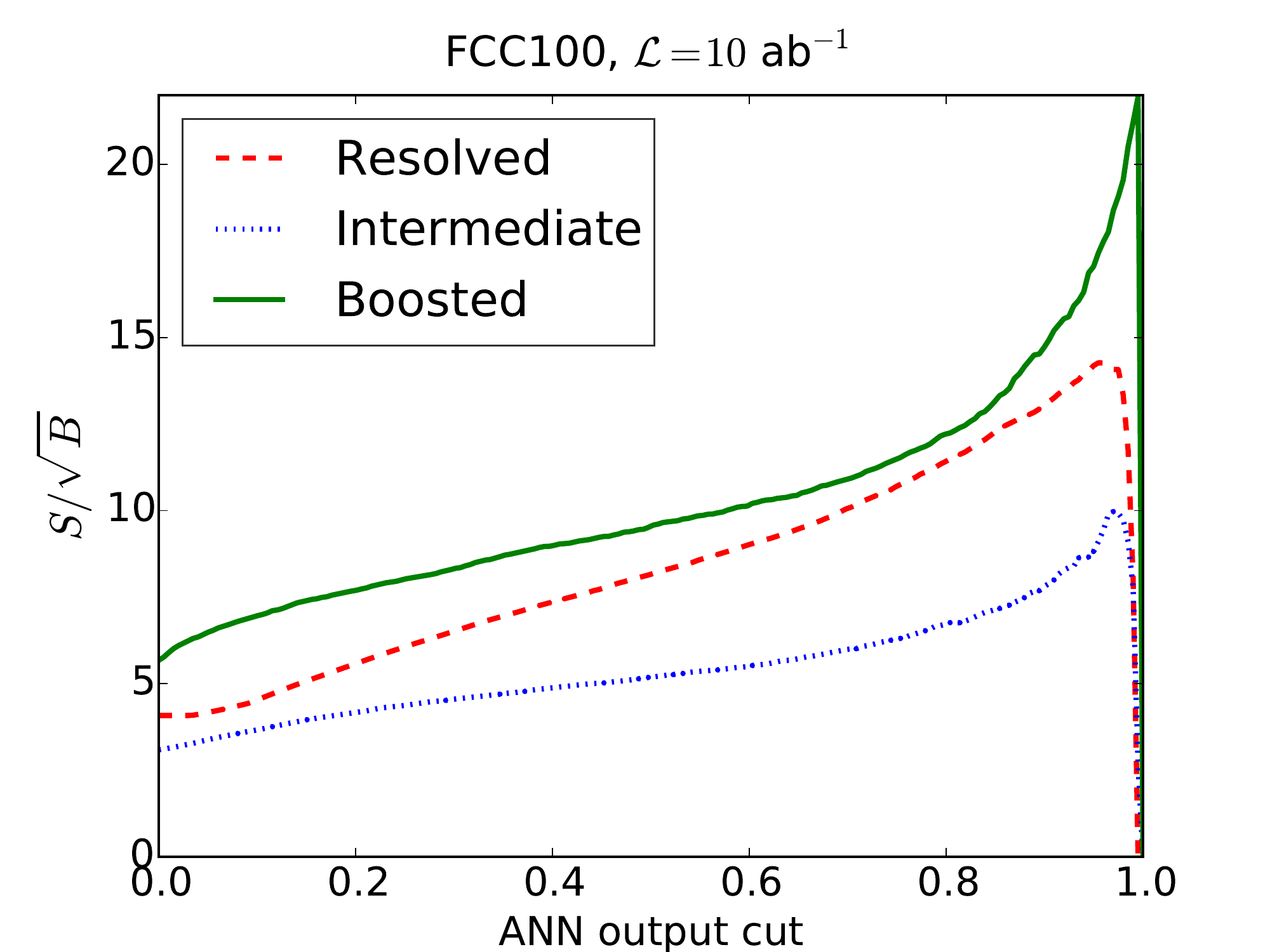}
\caption{\small Results of the neural network MVA analysis upon the LHC14 samples (upper plots) and FCC100 samples (lower plots). The left panels show the change in signal over background as the cut in neural network discriminator is varied. The right panels show the variation in statistical signal significance.}
\label{fig:mvares}
\end{center}
\end{figure}

Passing the results of the cut-based analyses through the neural network MVA, significant gains in both signal significance and $S/B$ are readily achievable. Figure~\ref{fig:mvares} demonstrates these gains for both the HL-LHC and FCC configurations once again including only the irreducible QCD background in both for comparison. The overall picture apparent after the cut-based analysis is retained here, with the FCC able to demonstrate extremely good signal significances but poorer signal to background ratios, with $S/B \sim 1\%$ achievable only for very aggressive cuts in the neural network discriminator. Such aggressive cuts could potentially yield signal significances of order 10 standard deviations or above, with the boosted topology at the FCC particularly enjoying the benefits of an increased centre of mass energy. In comparison, the HL-LHC results demonstrate poorer signal significance, although they are able to benefit from better than percentage level $S/B$ even for moderate cuts in the neural network output. Our analysis therefore suggesting that the HL-LHC environment has a significant competitive edge over an FCC when it comes to the crucial sensitivity to systematic error.

\section{Conclusion}
In this contribution we have examined how the prospects of measuring double Higgs production may change in a potential 100 TeV future circular collider with respect to what is acheivable at the high luminosity LHC. Using an analysis procedure efficient over a large range of scales, and a neural network based multivariate analysis, we are able to conclude that while the potential signal significances are likely to be much larger at an FCC, such a measurement would be plagued by an extremely poor signal to background ratio and therefore be damagingly sensitive to systematic uncertainties on the background. We therefore conclude that our current best prospects for studying SM double Higgs production and associated quantities such as the trilinear coupling are strongest at the HL-LHC.
\section{Acknowledgements}
Much of these studies were performed in the context of the FCC working group~\cite{Contino:2016spe}, and have been performed with the support of an ERC starting grant "PDF4BSM". 


\end{document}